\documentclass[seceq]{ptptex}

\usepackage{graphicx}
\usepackage{wrapft}

\title{%        %You can use \\ for explicit line-break
Charge screening effect in the hadron-quark mixed phase
}

\author{Tomoki \textsc{Endo}$^1$%       %Use \scshape  for the family name
, Toshiki \textsc{Maruyama}$^2$%
, Satoshi \textsc{Chiba}$^2$%
 and Toshitaka \textsc{Tatsumi}$^1$%
}

\inst{%         %Affiliation, neglected when [addenda] or [errata]
$^1$Department of Physics, Kyoto University, Kyoto 606-8502, Japan \\
$^2$Advanced Science Research Center, Japan Atomic Energy Research Institute,
 Tokai, Ibaraki 319-1195, Japan
}

\abst{%
The Coulomb interaction effect and the surface effect are consistently 
taken into account 
 in the hadron-quark mixed phase. These two finite-size effects
 greatly change the properties of the mixed phase and 
 restrict its density region.  In particular, the charge
 screening effect and the rearrangement of the charged particles are
 elucidated. Keeping the Gibbs conditions throughout the numerical
 procedure, we show
 the Maxwell construction effectively regain the physical meaning and the
 equation of state becomes similar to that given by the Maxwell construction. 
}

\begin{document}

\maketitle

\section{Introduction}\label{intro}

Deconfinement phase transition is believed to occur in hot and/or
high-density matter, while its mechanism has not been well understood 
yet. Many
authors have studied this transition by model
calculations or first-principle calculations like lattice QCD\cite{rev}.
Nowadays it is widely accepted that quark matter exists in
hot and/or high-density region like inner core of neutron stars. 
Static and dynamic properties of quark matter have been 
 actively studied theoretically for quark-gluon
 plasma (QGP), color superconductivity \cite{alf1,alf3} or magnetism
 \cite{tat1,tat2,tat3}.
Phenomenologically quark matter has been actively searched for in
relativistic heavy-ion collisions (RHIC)\cite{rhic1}, 
or in early universe and compact stars\cite{mad3,chen}. 

Many theoretical calculations have suggested that the deconfinement phase
transition should be of first order
in low temperature and high density area\cite{pisa,latt}.  
Therefore we assume first-order phase transition in this study. We,
hereafter, consider the phase transition from nuclear matter to
three-flavor quark matter in neutron-star matter for simplicity. If the
deconfinement transition is of first order, we may expect the {\it mixed
phase} during the transition.
The hadron-quark mixed phase has been considered during the
hadronization in RHIC\cite{rhic21,rhic22,rhic23}
or the boundary between quark matter and hadron matter in neutron
stars\cite{gle2}. 

There is an issue about the mixed phase for the first-order phase
transitions with more than one chemical potential\cite{gle1}.
We often use the Maxwell construction (MC) to derive the equation
of state (EOS) in thermodynamic equilibrium, as in the water-vapor phase
transition. In this case both phases consist of single particle species (${\rm H_2O}$).
However, if many particle species participate in the phase
transition as in neutron-star matter, MC is no more an appropriate
method. 
Before Glendenning first pointed out \cite{gle1}, 
many people have applied MC to get EOS of the  
first-order phase transitions\cite{weis,migd,elli,rose} expected in
neutron stars, such
as pion or kaon condensation and the deconfinement transition. 

For the deconfinement transition in neutron-star matter, 
we consider quark degrees of freedom
as well as hadrons and leptons. Accordingly we must introduce many chemical
potentials for particle species, but the independent ones
in this case
are reduced to two,
i.e.\ baryon-number chemical potential $\mu_\mathrm{B}$ and charge chemical potential $\mu_\mathrm{Q}$,
due to beta-equilibrium and total charge neutrality.
They are nothing but the neutron and electron chemical potentials, $\mu_n$ and
$\mu_e$, 
respectively. In the mixed-phase these chemical potentials should be
 spatially
constant. When we naively apply MC to get EOS in thermodynamic equilibrium, we
immediately notice that $\mu_\mathrm{B}$ is constant in the mixed-phase, while
$\mu_\mathrm{e}$ is different in each phase because of the difference of the electron number in
these phases. This is because MC uses EOS of bulk matter in each phase,
which is of locally
charge-neutral and uniform matter; many electrons are needed in hadron
matter to cancel the positive charge of protons,
while in quark matter total charge
neutrality is almost fulfilled without electrons.
Thus
\begin{equation}
\mu_{\mathrm{B}}^{\mathrm{Q}} = \mu_{\mathrm{B}}^{\mathrm{H}}, \hspace{10pt}
\mu_{\mathrm{e}}^{\mathrm{Q}} \neq  \mu_{\mathrm{e}}^{\mathrm{H}},
\label{chemeqMC}
\end{equation}
in MC, where superscripts ``Q'' and ``H'' denote the quark and hadron phase,
 respectively.
Glendenning emphasized that we must use the Gibbs conditions (GC)
in this case instead of MC, which relaxes the charge-neutrality
condition to be globally satisfied as a whole, not locally in each phase
\cite{gle1}. GC imposes the following conditions,
\begin{eqnarray}
 \mu_{\mathrm{B}}^{\mathrm{Q}} &= \mu_{\mathrm{B}}^{\mathrm{H}},
 \hspace{5pt} \mu_{\mathrm{e}}^{\mathrm{Q}} =
 \mu_{\mathrm{e}}^{\mathrm{H}}, \nonumber \\
P^{\mathrm{Q}} &=P^{\mathrm{H}} , \hspace{5pt} T^{\mathrm{Q}} = T^{\mathrm{H}}.
\end{eqnarray}
He demonstrated a wide region of the mixed phase, where two phases
have a net charge but totally charge-neutral:
EOS thus obtained,
different from
that given by MC,
never exhibits a constant-pressure region.
He simply considered
the mixed phase consisting of two bulk matters separated by a sharp
boundary without any surface
tension and the Coulomb interaction, which we call ``bulk Gibbs'' for convenience.

\begin{figure}[h!]\begin{center}
%\includegraphics[width=8cm]{edens_rho.eps}
%\begin{center}
\includegraphics[width=70mm]{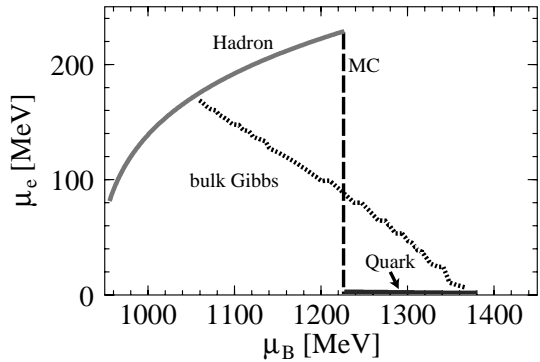}
\caption{ Phase diagram in the $\mu_\mathrm{B}-\mu_\mathrm{e}$ plane. 
There appears no region of the mixed phase by the calculation with MC, while a wide region of the mixed
		   phase by the ``bulk Gibbs'' calculation.}
\label{bulk}
\end{center}
\end{figure}

``Bulk Gibbs'' requires that each matter can have a net charge but total charge
is neutral, 
\begin{equation}
 f_V \rho_{\mathrm{ch}}^\mathrm{Q} + (1-f_V) \rho_{\mathrm{ch}}^\mathrm{H} = 0,
\end{equation}
where $f_V$ means the volume fraction of quark matter in the mixed phase
and ``$\rho_\mathrm{ch}^\mathrm{Q,H}$'' means charge density in each matter.
Figure \ref{bulk} shows the phase diagram in the $\mu_\mathrm{B}$ - $\mu_e$ plane.
We can see that there is a discontinuous jump in $\mu_e$ for the case of
MC, while the curve given by ``bulk Gibbs'' smoothly connects uniform hadron matter
and uniform quark matter; the mixed phase can appear in the wide
$\mu_\mathrm{B}$ region in ``bulk Gibbs'', in contrast with MC\cite{alf2}.
%\begin{wrapfigure}{l}{70mm}
%  \epsfxsize = \halftext
%  \centerline{ \epsfbox{edens_rho.eps}}
%\end{wrapfigure}

 However, this ``bulk Gibbs'' should be too simple to study the mixed phase,
 since we must consider non-uniform structures to be more realistic,  
instead of two bulk uniform matters; the mixed phase should have
 various geometrical structures, where both the number and charge
 densities are no more uniform. 
Then we have to take into account the
 finite-size effects like the surface and the Coulomb interaction energies.
In this paper we study the structured mixed phase by treating the
 finite-size effects self-consistently. 
We show that the mixed phase should be narrow in
 the $\mu_\mathrm{B}$ space by the charge
 screening effect, and derive EOS for the deconfinement transition in
 neutron-star matter. We shall
 see it results in EOS being similar to the one given by MC. We also discuss the
 interplay of the Coulomb interaction effect and the surface effect in the context
 of the hadron-quark mixed phase.
Preliminary results for the droplet case has been already
reported in Ref.\cite{end1}.

The plan of the paper is as follows.
We briefly review the previous works in Sec.\ \ref{history}.
Section \ref{formalism} is devoted to our formalism and the numerical procedure.
We show the results of our calculation and
discuss the screening effect in Sec.\ \ref{results}. 
Finally, a summary and concluding remarks are given in Sec.\ \ref{summary}.

\section{Brief review of the previous works}\label{history}

 Heiselberg et al.\ \cite{pet} studied a geometrical structure in the mixed
 phase: they considered spherical quark droplets embedded in hadron
 matter by including the surface and the Coulomb energies. 
 They introduced the surface tension and treated its strength as a free parameter because the
 surface tension at the hadron-quark interface has not been clearly
 understood. They
 pointed out that if the surface tension parameter $\sigma$ is large
 ($\sigma \geq$ 90 MeV/fm$^2$), the region of the mixed phase is largely
 limited  or cannot exist. 
Subsequently 
Glendenning and Pei \cite{gle2} have suggested the ``crystalline structures
of the mixed phase'' which have 
some geometrical structures, ``droplet'', ``rod'', ``slab'', ``tube'', and ``bubble'',
assuming a small $\sigma$\cite{gle2,gle3}.

The finite-size effects are obvious in these 
 calculations by observing energies.
We may consider only a single cell, by dividing the whole space into 
equivalent Wigner-Seitz cells: the cell size is denoted by $R_W$ and the
size of the lump (droplet, rod, slab, tube or 
bubble) by $R$. Then the surface energy density 
is expressed in terms of the surface tension parameter $\sigma$ as 
\begin{eqnarray}
\epsilon_\mathrm{S} = \frac{f_V \sigma d}{R},
\end{eqnarray}
where $d$ denotes the
dimensionality of each geometrical structure; 
$d=3$ for droplet and bubble,
$d=2$ for rod and tube, and $d=1$ for slab. 
The Coulomb energy density reads
\begin{eqnarray}
\epsilon_\mathrm{C} &=& 2\pi e^2 \left(\rho_\mathrm{ch}^\mathrm{H}
- \rho_\mathrm{ch}^\mathrm{Q}  \right)^2 R^2 \Phi_d(f_V) \hspace{5pt},\\
\Phi_d(f_V) &\equiv& \left[2 (d-2)^{-1} \left( 1 -\frac{1}{2} d f_V^{1-2/d}  \right)
+ f_V  \right] \left(d+2 \right)^{-1},
\end{eqnarray}
where we simply assumed uniform density in each phase.
When we minimize the sum of $\epsilon_S$ and $\epsilon_C$ with respect
to the size $R$ for a given volume fraction $f_V$, we can get the well
known relation, 
\begin{equation}
\epsilon_S = 2 \epsilon_C.
\label{edensCS}
\end{equation}
This implies that an
optimal size of the lump is determined by the balance of these finite-size effects.
Eventually we can express
the size-dependent energy density besides the bulk energy density \cite{pet}:
\begin{equation} 
\epsilon^{(d)}_\mathrm{C} + \epsilon^{(d)}_\mathrm{S} = 3 f_V d \left(
							      \frac{\pi
							      \sigma^2
							      \left(
							       \rho_\mathrm{ch}^\mathrm{H} - \rho_\mathrm{ch}^\mathrm{Q}  \right)^2 \Phi_d (f_V)}{2d}  \right)^{1/3}. 
\label{bulk_cs} 
\end{equation}
Thus we can calculate the energy of any geometrically structured mixed phase with
(\ref{bulk_cs}) by changing the parameter $d$. 
Many authors have taken this treatment for the mixed
phase\cite{gle2,alf2,pet}.
Note that the energy sum in Eq.\ (\ref{bulk_cs}) becomes larger as the
surface tension gets stronger, while the relation  Eq.(\ref{edensCS}) is always kept.

However, this treatment is not a self-consistent, but a perturbative one,
since the charge screening effect for the Coulomb potential or the rearrangement of
charged-particles in the presence of the Coulomb interaction is completely discarded. 
We shall see that the Coulomb potential is never weak in the mixed
phase, and thereby this treatment overestimates the Coulomb energy.
The charge screening effect is included only if we introduce 
the Coulomb potential and consistently
solve the Poisson equation with other equations of motion for charged
particles. Consequently it is a highly non-perturbative effect.
Norsen and Reddy\cite{nors} have studied the Debye screening effect in
the context of kaon condensation to see a large change of the
charged-particle densities like kaons and protons. 
Maruyama et al.\ have numerically studied it in the context of liquid-gas phase
transition at subnuclear densities \cite{maru1}, where nuclear pastas can be
regarded as geometrical structures in the mixed phase. Subsequently,
they have also studied kaon
condensation at high-densities \cite{maru2}, where we have seen that kaonic pastas appear
in the mixed phase. Through these works we have figured out the role of
the Debye screening in the mixed phase. We have also studied the
interplay of the Coulomb effect and the surface effect. 

Voskresensky et al.\ \cite{vos} explicitly  studied the Debye screening
 effect for a few geometrical structures of the hadron-quark mixed
 phase.
 They have shown that
 the optimal value of the size of the structure cannot be obtained
 due to the charge screening
 even if the surface tension is not so strong.
 They called it as mechanical instability.
 It occurs because the Coulomb
 energy density is suppressed at larger size than the Debye screening length
 (cf.\ Eq.~(\ref{lambda})).
They also suggested that the properties of the mixed phase become very
 similar to those given by MC, if the charge screening effectively
 works.
They also noted that the apparent violation of the Gibbs condition (Eq.~(\ref{chemeqMC}))
 can be remedied by including the Coulomb potential in a
 gauge-invariant way: the number of the charged particles is given by a
 gauge invariant combination of the chemical potential and the Coulomb
 potential, and thereby the number can be different in each phase for a
 constant charge chemical potential if the 
 the Coulomb potential takes different values in both phases.
However, they used a linear approximation to solve the Poisson equation analytically.

If the Coulomb interaction effect is so important, it would be important
to study it without
recourse to any approximation.
In this paper we numerically study the charge screening effect on the
structured 
mixed phase during the deconfinement transition in neutron-star matter  
in a self-consistent way.
Actually we shall see importance of non-linear effects included in the
Poisson equation. 

\section{Self-consistent calculation}\label{formalism}

\subsection{Thermodynamic potential}

 We consider the geometrically structured mixed phase (SMP) where one
 phase is embedded in the other phase with a certain geometrical form.
%
%\begin{figure}[htb]
%\begin{minipage}[t]{80mm}
\begin{wrapfigure}{l}{50mm}
%  \epsfxsize = \halftext
%  \centerline{ \epsfbox{edens_rho.eps}}
%\includegraphics[width=8cm]{edens_rho.eps}
\begin{center}
\includegraphics[width=30mm,height=60mm,keepaspectratio]{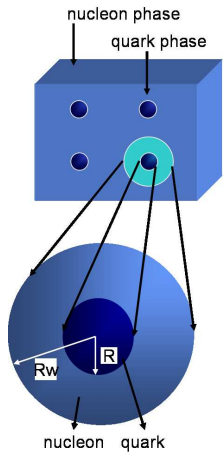}
\caption{Wigner-Seitz approximation for the droplet case. $R$ is the
 droplet radius and $R_\mathrm{W}$ the cell radius.}
\label{wsapp}
%\end{minipage}
\end{center}
%\end{figure}
\end{wrapfigure}
We divide the whole space into equivalent charge-neutral Wigner-Seitz cells
with a size $R_W$  and a size of embedded phase $R$  as illustrated in Fig.\ \ref{wsapp}.
Quark phase consists of {\it u}, {\it d}, {\it s} quarks and
electron.  Hadron phase consists of proton, neutron and electron.
We incorporate the MIT Bag model and assume the sharp boundary at the
hadron-quark interface.
We use density functional
 theory (DFT) and incorporate local density approximation (LDA)
 \cite{parr,drez}. 

We consider total thermodynamic potential ($\Omega_\mathrm{total}$) 
which consists of the hadron, quark and electron and the Coulomb
interaction contributions:
\begin{equation}
\Omega_\mathrm{total} = \Omega_\mathrm{hadron} +\Omega_\mathrm{quark} +\Omega_\mathrm{em},
\label{ometot}
\end{equation}
where we summarize the contributions of electrons and the Coulomb
interaction 
as $\Omega_\mathrm{em}$
because they are present in both phases.

We briefly present the expressions of
 thermodynamic potentials. The details of the derivation of the expressions are
 given in Ref.\cite{end2}.

First, the Coulomb interaction energy is expressed in terms of
particle densities,
\begin{equation}
E_V = \frac{1}{2} \sum_{i,j} \int d^3 r d^3 r^{\prime} \frac{Q_i \rho_i(\vec{r}) Q_j \rho_j(\vec{r}^{\prime})}{\left| \vec{r} - \vec{r}^{\prime} \right|},
\end{equation}
where $i=u,d,s,p,n,e$ with $Q_i$ being the particle charge ($Q=-e < 0$ for the electron).
Accordingly the Coulomb potential is defined as
\begin{equation}
V (\vec{r}) = -\sum_i \int d^3 r^{\prime} \frac{e Q_i
 \rho_i(\vec{r}^{\prime})}{\left| \vec{r} - \vec{r}^{\prime} \right|}+V_0,
\label{vcoul}
\end{equation}
where $V_0$ is an arbitrary constant representing the gauge degree of
 freedom. 
We fix the gauge by a condition $V(R_W) = 0$ in this paper (see Sec.\ 2.2). 
Operating a Laplacian $\nabla^2$ on the Coulomb potential
$V(\vec{r})$, we automatically derive the Poisson equation.

Therefore, the electron contribution and the Coulomb interaction energy (in both phases) are expressed as
\begin{eqnarray}
\Omega_{\mathrm{em}} \! \! &=& \! \! \int \! d\vec{r} \Biggl[  -\frac{1}{8\pi e^2} \left(
							       \nabla V
							       (\vec{r})
							      \right)^2 
+ \epsilon_e (\rho_e (\vec{r})) - \mu_e \rho_e (\vec{r}) +  V (\vec{r}) \rho_e (\vec{r})  \Biggr] \nonumber \\
 \! \!  &=& \! \int \! d\vec{r} \left[ -\frac{1}{8\pi e^2} \left( \nabla V (\vec{r})
					      \right)^2 \!-\! \frac{\left(V
					      (\vec{r}) \!- \! \mu_e \right)^4}{12\pi^2} \! \right], 
\end{eqnarray} 
where
$\epsilon_e (\rho_e(\vec{r}))=\frac{\left( 3 \pi^2 \rho_e (\vec{r})\right)^\frac{4}{3}}{4\pi^2}$
 is the kinetic energy density of electron.

 Secondly, in the quark phase, {\it u} and {\it d}
 quarks are treated as massless particles and only {\it s} quark massive
 one, $m_s= 150$ MeV.
The kinetic energy of quark of flavor $f$ is simply expressed as\cite{tama}
\begin{equation}
\epsilon_{f \mathrm{kin}}= \frac{3}{8 \pi^2} m_f^4 \left[ x_f \eta_f \left( 2x_f^2+1 \right) -\ln\left( x_f +\eta_f  \right) \right],
\end{equation}
where $m_f$ is the quark mass, $x_f = p_{\mathrm{F}f}(\vec{r})/m_f$ with
Fermi momentum $p_{\mathrm{F}f}(\vec{r}) = (\pi^2 \rho_f (\vec{r}) )^\frac{1}{3}$ and $\eta_f=\sqrt{1+x^2_f}$.

For the interaction energy, we take into account the leading order
contribution coming from the one-gluon exchange interaction. 
Since the contribution from the Hartree term disappears due to the traceless
 property of the Gell-Mann matrix ($\lambda$),
 the leading order contribution only comes from the Fock term, 
\begin{equation}
 \epsilon_{f \mathrm{Fock}} = -\frac{\alpha_{\mathrm{c}}}{\pi^3}  m_f^4 \left\{ x_f^4 - \frac{3}{2} \left[ x_f \eta_f -\ln\left( x_f +\eta_f  \right)  \right]^2  \right\}. 
\label{fock}
\end{equation}
Including this interaction, the quark contribution to the thermodynamic potential is expressed as
\begin{eqnarray}
& & \Omega_{\mathrm{quark}} =
 \Omega_{\mathrm{u}}+\Omega_{\mathrm{d}}+\Omega_{\mathrm{s}} +\int d\vec{r} B , \label{omeq}\\
\Omega_f  &=& \int \! d\vec{r} \left[
\epsilon_f (\rho_f (\vec{r})) - \mu_f \rho_f (\vec{r}) -
				    N_i V (\vec{r})
				    \rho_f (\vec{r}) \right] , \hspace{10pt} N_i=\frac{Q_i}{e},
\nonumber 
\end{eqnarray}
where, the energy density $\epsilon_f (\rho_f
(\vec{r}))$ stands for  $\epsilon_{f \mathrm{kin}} + \epsilon_{f
\mathrm{Fock}}$ of $f$ quark, and $B$ is the Bag constant.
The Bag constant is taken as 120 MeV/fm$^3$, and the QCD fine structure constant 
 as $\alpha_{\mathrm{c}}=0.4$, which are also used by Heiselberg et
 al.\ \cite{pet} and in the previous work\cite{vos}.

Thirdly, we consider the hadron contribution. 
The thermodynamic potential for the non-relativistic nucleons becomes
\begin{eqnarray}
\Omega_{\mathrm{hadron}} = E_N - \sum_{a=p, n} \mu_a \int d\vec{r} \rho_a
 (\vec{r}) -  \int d \vec{r} \, V (\vec{r}) \rho_{\mathrm{p}}(\vec{r}) ,
\label{omeh}
\end{eqnarray}
where $E_N$ is the energy of the nucleons,
\begin{eqnarray}
E_N = \int d \vec{r} \left[ \sum_{a=p, n} \frac{3}{10m}\left( 3 \pi^2 \right)^\frac{2}{3} \rho^\frac{5}{3}_a
 (\vec{r}) + \epsilon_\mathrm{pot}  \left( \rho_{\mathrm{p}} (\vec{r})
				      , \rho_{\mathrm{n}} (\vec{r})
				     \right) \right].
\end{eqnarray}
Here we use the effective potential $\epsilon_\mathrm{pot} (\rho_{\mathrm{p}} (\vec{r}), \rho_{\mathrm{n}}
(\vec{r}))$ parametrized by the local densities for simplicity,
\begin{eqnarray}
\epsilon_\mathrm{pot} (\vec{r})  &=& S_0 \frac{\left( \rho_\mathrm{n}(\vec{r}) - \rho_\mathrm{p}(\vec{r})  \right)}{\rho_0(\vec{r})} + \left( \rho_\mathrm{n}(\vec{r}) + \rho_\mathrm{p}(\vec{r})  \right) \epsilon_\mathrm{bind} \nonumber \\
&+&  K_0  \frac{\left( \rho_\mathrm{n}(\vec{r}) + \rho_\mathrm{p}(\vec{r})  \right)}{18} \left( \frac{\rho_\mathrm{n}(\vec{r}) + \rho_\mathrm{p}}{\rho_0} - 1  \right)^2 \nonumber\\
&+&  C_\mathrm{sat}  \left( \rho_\mathrm{n}(\vec{r}) +
			     \rho_\mathrm{p}(\vec{r})  \right)  \left(
			     \frac{\rho_\mathrm{n}(\vec{r}) +
			     \rho_\mathrm{p}(\vec{r})}{\rho_0}  -  1
			      \right) ,
\label{effpot}
\end{eqnarray}
where $S_0$,  $K_0$,  $\epsilon_\mathrm{bind}$, and $C_\mathrm{sat}$ are adjustable parameters 
to reproduce the saturation properties of nuclear
matter\cite{vos}. 
 We consider beta equilibrium at the hadron-quark
 interface as well as in each phase:
\begin{eqnarray}
&& \mu_\mathrm{u}+\mu_\mathrm{e} = \mu_\mathrm{d},\nonumber\\
&& \mu_\mathrm{d} = \mu_\mathrm{s}, \nonumber \\
&& \mu_\mathrm{p}+\mu_\mathrm{e} = \mu_\mathrm{n} \equiv \mu_\mathrm{B}, \nonumber \\
&& \mu_\mathrm{n} = \mu_\mathrm{u}+2 \mu_\mathrm{d}, \nonumber\\
&& \mu_\mathrm{p} = 2 \mu_\mathrm{u}+\mu_\mathrm{d}.
\label{chemeq}
\end{eqnarray}
The last relation can be derived from other four relations,
so that there are left four independent conditions for chemical equilibrium.

We get the equations of motion from
$\frac{\delta \Omega_\mathrm{total}}{\delta\phi_i}=0$ 
( $\phi_{i}=\rho_u(\vec{r}), \rho_d(\vec{r}), \rho_s(\vec{r}),$ 
$\rho_p(\vec{r}), \rho_n(\vec{r}), \rho_e(\vec{r}), V(\vec{r})$ ):
The Poisson equation then reads
\begin{eqnarray}
\nabla^2 V (\vec{r}) \! \! = \! \! 4 \pi e^2 \left[ \left(\frac{2}{3}\rho_u
					 (\vec{r})- \frac{1}{3}\rho_d
					 (\vec{r}) - \frac{1}{3} \rho_s
					 (\vec{r}) \right)
+ \rho_{\mathrm{p}} (\vec{r}) -  \rho_e (\vec{r})  \right].
\label{poisson}
\end{eqnarray}
Other equations of motion give nothing but the expressions 
of the chemical potentials,
\begin{equation}
\mu_i = \frac{\delta E_\mathrm{kin+str}}{\delta \rho_i (\vec{r})} - N_i V(\vec{r}),
\label{mu_i}
\end{equation}
where $\displaystyle E_\mathrm{kin+str} = \sum_{i= u, d,
s,e} \int d\vec{r} \epsilon_i + E_N$.
Then quark chemical potentials are expressed as
\begin{eqnarray}
\mu_{\mathrm{u}} &=& \left( 1 + \frac{2 \alpha_{\mathrm{c}}}{3 \pi}  \right) \pi^\frac{2}{3} \rho_{\mathrm{u}}^\frac{1}{3} (\vec{r})- \frac{2}{3} V (\vec{r})\\
\mu_{\mathrm{d}} &=& \left( 1 + \frac{2 \alpha_{\mathrm{c}}}{3 \pi}  \right) \pi^\frac{2}{3} \rho_{\mathrm{d}}^\frac{1}{3}(\vec{r}) + \frac{1}{3} V (\vec{r})\\
\mu_{\mathrm{s}} &=& \epsilon_{{\mathrm{Fs}}}(\vec{r}) + \frac{2
 \alpha_{\mathrm{c}}}{3 \pi} \left[ p_{\mathrm{Fs}}(\vec{r})- 3
			      \frac{m_{\mathrm{s}}^2}{\epsilon_{\mathrm{Fs}}(\vec{r})} \ln \left( \frac{\epsilon_{\mathrm{Fs}}(\vec{r})+p_{\mathrm{Fs}}(\vec{r})}{m_{\mathrm{s}}} \right)  \right]  + \frac{1}{3} V (\vec{r}), \nonumber \\
\end{eqnarray}
with $\epsilon_{\mathrm{Fs}} (\vec{r})=
\sqrt{m_{\mathrm{s}}^2+p_{\mathrm{Fs}}^2(\vec{r})}$.
On the other hand chemical potentials of nucleons and electrons are
\begin{eqnarray}
\mu_{\mathrm{n}} &=& \frac{p_{\mathrm{Fn}}^2}{2m} + \frac{2S_0 \left(\rho_n(\vec{r})-\rho_{\mathrm{p}}(\vec{r})
\right)}{\rho_0}+ \epsilon_{\mathrm{bind}} \nonumber \\
&+& \frac{K_0}{6} \left( \frac{\rho_{\mathrm{n}} (\vec{r}) \!+\!
		 \rho_{\mathrm{p}} (\vec{r})} {\rho_0} - 1  \right)^2 +
\frac{K_0}{9} \left(  \frac{\rho_{\mathrm{n}} (\vec{r})+
	       \rho_{\mathrm{p}} (\vec{r})}{\rho_0}- 1  \right) \nonumber \\
&+& 2 C_{\mathrm{sat}}  \frac{\rho_{\mathrm{n}}(\vec{r}) + \rho_{\mathrm{p}}(\vec{r})}{\rho_0} - C_{\mathrm{sat}} \\
\mu_{\mathrm{p}} &=& \mu_{\mathrm{n}} - \frac{p_{\mathrm{Fn}}^2
 (\vec{r})}{2m}+ \frac{p_{\mathrm{Fp}}^2 (\vec{r})}{2m} - \frac{4 S_0
 \left( \rho_{\mathrm{B}} - 2 \rho_{\mathrm{p}} (\vec{r})
 \right)^2}{\rho_0} - V (\vec{r}) \nonumber \\ 
\mu_e &=& \left( 3 \pi^2 \rho_e (\vec{r}) \right)^\frac{1}{3} + V (\vec{r}).
\end{eqnarray}
We solve these equations of motion under GC. 
Important point is that the Coulomb potential $V(\vec{r})$ 
is included in each expression in a proper way.
The Coulomb potential is the
function of charged-particle densities, and in turn
densities are functions of the Coulomb potential. As a result, the Poisson
equation becomes highly non-linear. Since it should be difficult to solve them
analytically, we numerically solve them without any approximation. 

Once the geometrical structure is concerned,
we have to take into account the surface tension as well 
at the interface of the hadron and quark phases. 
It may be connected with the confining mechanism and unfortunately 
we have no definite idea about how to incorporate it. 
Actually many authors
have treated its strength as a free parameter and seen 
how the results are changed by its value\cite{gle2,pet,alf2}. 
Here we also follow this manner by introducing the surface tension
parameter $\sigma$ to simulate the surface effect.
One might be afraid that the surface tension will be modified 
once the Coulomb interaction is explicitly introduced. 
However, such modification might be rather small, 
as inferred from the previous result\cite{maru2}. 

 Note that we have to determine now eight variables, i.e., six chemical potentials, $\mu_\mathrm{u}$, $\mu_\mathrm{d}$,
 $\mu_\mathrm{s}$, $\mu_\mathrm{p}$, $\mu_\mathrm{n}$, 
$\mu_\mathrm{e}$, and  radii $R$ and $R_W$.
First, we fix $R$ and $R_W$.
Here we have four conditions due to
 $\beta$ equilibrium (\ref{chemeq}).
Therefore, once two chemical potentials $\mu_\mathrm{B}$ and
$\mu_\mathrm{e}$ are given, 
we can determine other four chemical potentials, 
$\mu_\mathrm{u}$, $\mu_\mathrm{d}$, $\mu_\mathrm{s}$ and $\mu_\mathrm{p}$. 
Next, we determine $\mu_\mathrm{e}$ by the global charge neutrality
condition: 
\begin{equation}
 f_V \rho_{\mathrm{ch}}^{\mathrm{Q}} + (1-f_V) \rho_{\mathrm{ch}}^{\mathrm{H}} = 0,
\end{equation}
where the volume fraction $f_V=\left(\frac{R}{R_W}\right)^d$, and $d$
denotes the dimensionality of each geometrical structure.
At this point $f_V$ is still fixed.

The pressure coming from the surface tension is given by 
\begin{equation}
P_{\sigma}= \sigma \frac{d S}{d V_{\rm Q}},
\end{equation}
 where $S$ is the area of the surface and $V_{\rm Q}$ is the volume of the
 quark phase.
Then we find the optimal value of $R$ ($R_W$ is fixed and 
thereby $f_V$ is changed by $R$) 
by using one of GC;
\begin{equation}
P^\mathrm{Q} = P^\mathrm{H} + P_\sigma.
\label{pbalance}
\end{equation}
The pressure in each phase $P^\mathrm{Q(H)}$ 
is given by the thermodynamic relation: $
P^\mathrm{Q(H)}=-\Omega_\mathrm{Q(H)}/V_\mathrm{Q(H)}$,
where $\Omega_\mathrm{Q(H)}$ is the thermodynamic potential in each
phase and given by adding electron and the Coulomb interaction
contributions to $\Omega_\mathrm{quak(hadron)}$ in Eqs.\ (\ref{omeq}) and (\ref{omeh}).
Finally, we determine $R_W$ by minimizing thermodynamic potential. 
Therefore once
$\mu_\mathrm{B}$ is given, all other values $\mu_i$ ($i=u,d,s,p,e$)
and $R$, $R_W$ can be obtained.

Note that we keep GC throughout the numerical procedure. 
We will see later how
the mixed phase would be changed by including finite-size effects 
keeping GC completely. 
Although MC is not rigidly
correct as we have seen in Sec.\ 1, our results will
show a similar behavior to those by MC as a result of including the finite-size effects.

In numerical calculation, every point inside the cell is represented by a grid point
(the number of grid points $N_\mathrm{grid} \approx 100 $). 
Equations of motion are solved by
a relaxation method for a given baryon-number chemical potential under
constraints of the global charge neutrality.

\subsection{Proper treatment of the Coulomb interaction}

With the Coulomb potential (\ref{vcoul}) and 
thermodynamic potentials (\ref{ometot}), 
the gauge invariance of our treatment can easily be seen as follows:
varying the expression of chemical potentials (\ref{mu_i}) with respect to
the Coulomb potential $V(\vec{r})$, as is shown in the previous work\cite{vos},
we have
\begin{equation}
A_{ij} \frac{\partial \rho_j }{\partial V } = N_i , \hspace{10pt}
\hspace{5pt} A_{ij} B_{jk} = \delta_{ik},
\end{equation}
where matrices $A$ and $B$ are defined as 
\begin{equation}
A_{ij} \equiv \frac{ \delta^2 E_{\mathrm{kin+str}}}{\delta \rho_i \delta
 \rho_j}
 \hspace{10pt} B_{ij} \equiv \frac{\partial \rho_i}{\partial \mu_j}.
\end{equation}
From these equations, the gauge-invariance relation follows,
\begin{equation}
\frac{\partial \rho_i}{\partial V} = N_j \frac{\partial \rho_j}{\partial \mu_i}.
\end{equation}
We can understand that chemical potential is gauge variant from this relation. 
When the Coulomb potential is
shifted by a constant value, $V(\vec{r}) \Longrightarrow V({\vec{r}}) - V_0$,
the charge chemical potential should be also shifted as $\mu_i \Longrightarrow
\mu_i+N_i V_0$. 
To take into account the Coulomb interaction, we have to include
$V(\vec{r})$ in the gauge invariant way like in Eq.\ (\ref{mu_i}). 
Note that the phase diagram in the $\mu_\mathrm{B}-\mu_\mathrm{e}$ plane (see ,e.g.,
Fig.~\ref{bulk}) is not well-defined,
since the charge chemical potential $\mu_e$ is not gauge invariant by
itself.

\section{Numerical results}\label{results}

\begin{figure}[htb]
\begin{minipage}[t]{70mm}
\includegraphics[width=70mm]{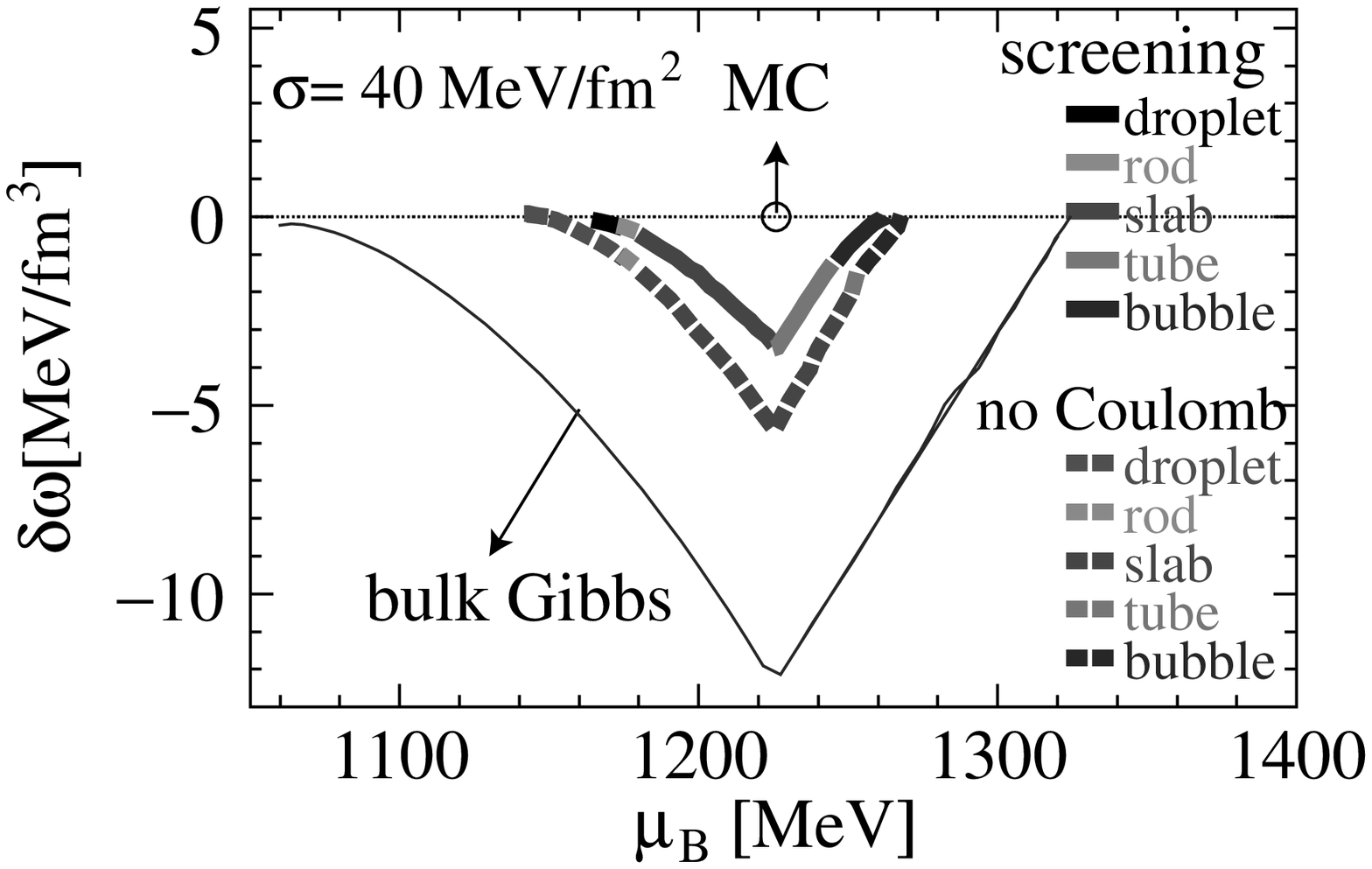}
\caption{ Difference of the thermodynamic potential density
 as a function of baryon-number chemical potential $\mu_\mathrm{B}$ for
 $\sigma=40$ MeV/fm$^2$. If $\delta \omega$ is negative, the mixed phase
 is a thermodynamically favorable state. MC determines one point of phase transition in uniform matter,
denoted as a circle in the $\mu_B$-$\delta\omega$ plane.}
\label{omed40}
\end{minipage}
\hspace{8pt}
\begin{minipage}[t]{70mm}
\includegraphics[width=70mm]{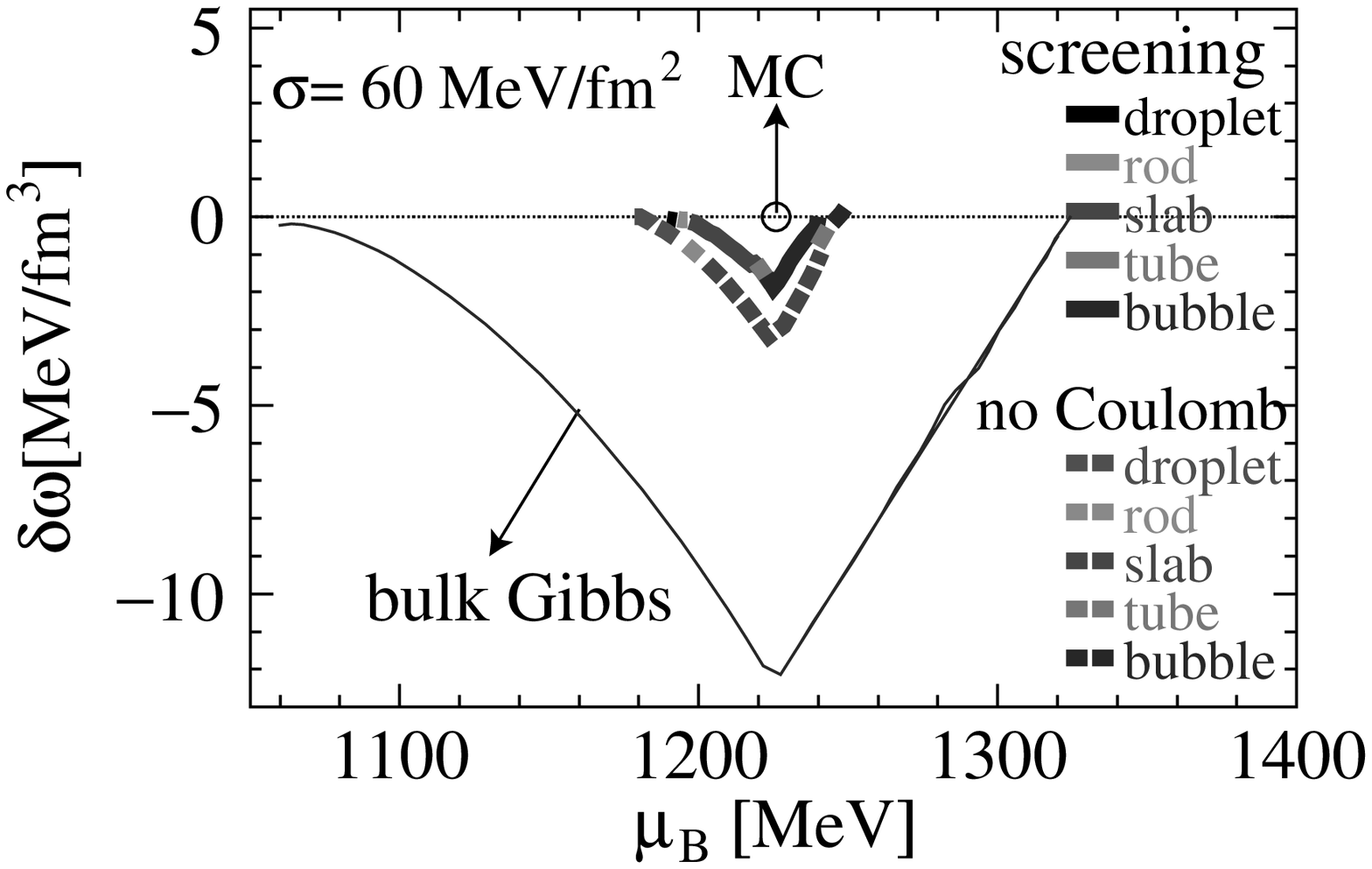}
\caption{ Same as Fig.~3 for $\sigma=60$ MeV/fm$^{2}$. The negative $\delta
 \omega$ region is narrower than the $\sigma=40$ MeV/fm$^{2}$ case.}
\label{omed60}
\end{minipage}
\end{figure}

We show the thermodynamic potential in Figs.\ \ref{omed40} and \ref{omed60}.
In uniform matter, hadron phase is thermodynamically favorable for 
$\mu_\mathrm{B} < 1225$ MeV and quark phase for $\mu_\mathrm{B} > 1225$.
Therefore
we plot $\delta \omega$, difference of the thermodynamic potential density
between the mixed phase and each uniform matter:
\begin{equation}
\delta \omega = \begin{cases} \omega_\mathrm{total}-\omega^\mathrm{uniform}_\mathrm{H} \quad \mu_\mathrm{B}
\leq 1225 \mathrm{MeV}, \\
\omega_\mathrm{total}-\omega^\mathrm{uniform}_\mathrm{Q} \quad \mu_\mathrm{B} \geq 1225 \mathrm{MeV},
\end{cases}
\end{equation}
where $\omega_\mathrm{total}=\Omega_\mathrm{total}/V_W$, etc.
There we also depict two results for comparison: 
one is given by the ``bulk Gibbs'' calculation, 
where the finite-size effects are completely discarded. 
The other is the thermodynamic potential given by a perturbative
treatment of the Coulomb interaction, which is denoted by ``no Coulomb''; 
discarding the Coulomb potential $V(\vec{r})$, 
we solve the equations of motion to get density profiles, 
then evaluate the Coulomb interaction energy by
using the density profiles thus determined. 
We can see the screening effects by comparing 
this ``no Coulomb'' calculation with the self-consistent one denoted by ``screening''.
$\delta \omega$ given by MC appears as a point denoted by a circle in
Figs.\ \ref{omed40} and \ref{omed60} where two conditions, $P^\mathrm{Q}=P^\mathrm{H}$ and
$\mu_\mathrm{B}^\mathrm{Q}=\mu_\mathrm{B}^\mathrm{H}$, are satisfied.
On the other hand the mixed phase derived from ``bulk Gibbs'' appears in a
wide region of $\mu_{\mathrm{B}}$. 
Therefore, if the region of the mixed phase becomes
narrower, it signals that the properties of the mixed phase become close
to those of MC. 
One may clearly see that $\omega_{\mathrm{total}}$ becomes close to that
given by MC due to the finite-size effects, the effects of the surface
tension and the Coulomb interaction.
The large increase of $\delta\omega$ from the ``bulk Gibbs'' curve comes from the effects
of the surface tension and the Coulomb potential. Since the surface
tension parameter is introduced by hand, we must carefully study the
effects of the surface tension and the Coulomb interaction, separately. 
 From the difference between the result
given by ``no Coulomb'' and that by ``bulk Gibbs'', we 
can roughly say that about $2/3$ of the increase comes from the
effect of the surface tension and $1/3$ from the Coulomb interaction
(see Eq. (\ref{edensCS})). 

Comparing the result of self-consistent calculation with that of ``no Coulomb'',
we can see that the change of energy 
caused by the screening effect is not so large, but still the same order of
magnitude as that given by the surface effect.
%
%\begin{wrapfigure}{r}{140mm}
%  \epsfxsize = \halftext
%  \centerline{ \epsfbox{edens_rho.eps}}
\begin{figure}[htb]
\begin{minipage}[t]{70mm}
%\includegraphics[width=8cm]{edens_rho.eps}
%\begin{center}
%\begin{tabular}{c c}
\includegraphics[width=70mm]{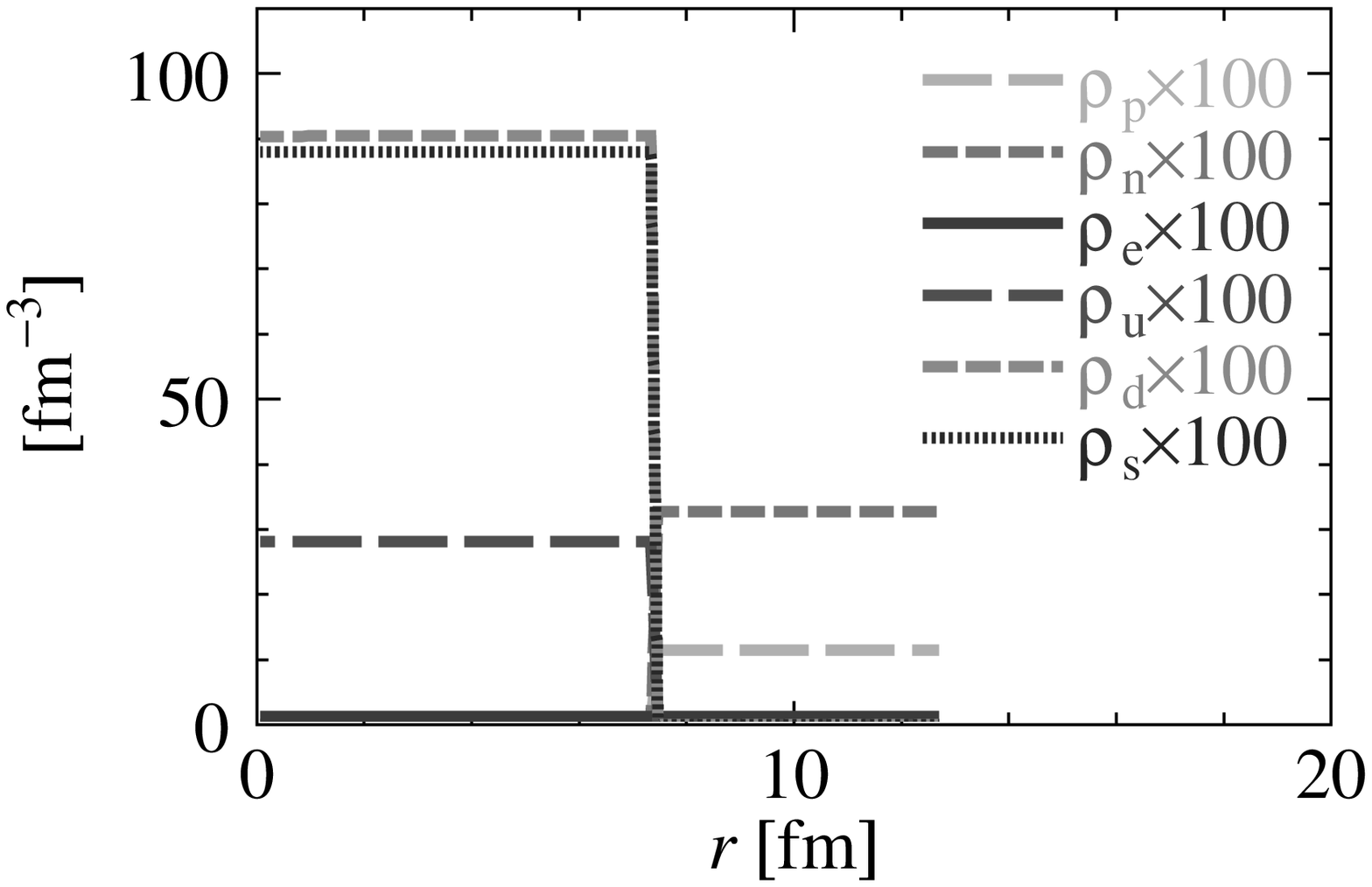}
\caption{ Density profiles in the droplet phase given by ``no Coulomb'' 
for $\mu_B=1189$ MeV and $\sigma=60$ MeV/fm$^{2}$. 
 They are uniform in each phase. 
$R=7.2$ fm and $R_W=12.8$ fm.}
\label{densprof-no}
\end{minipage}
%\hspace{8pt}
\begin{minipage}[t]{70mm}
%
%\caption{ Density profile for screening case for slab ($\mu_B=1218$ MeV,
% $\sigma=60$ MeV/fm$^{2}$). we can see density rearrangement.}
\includegraphics[width=70mm]{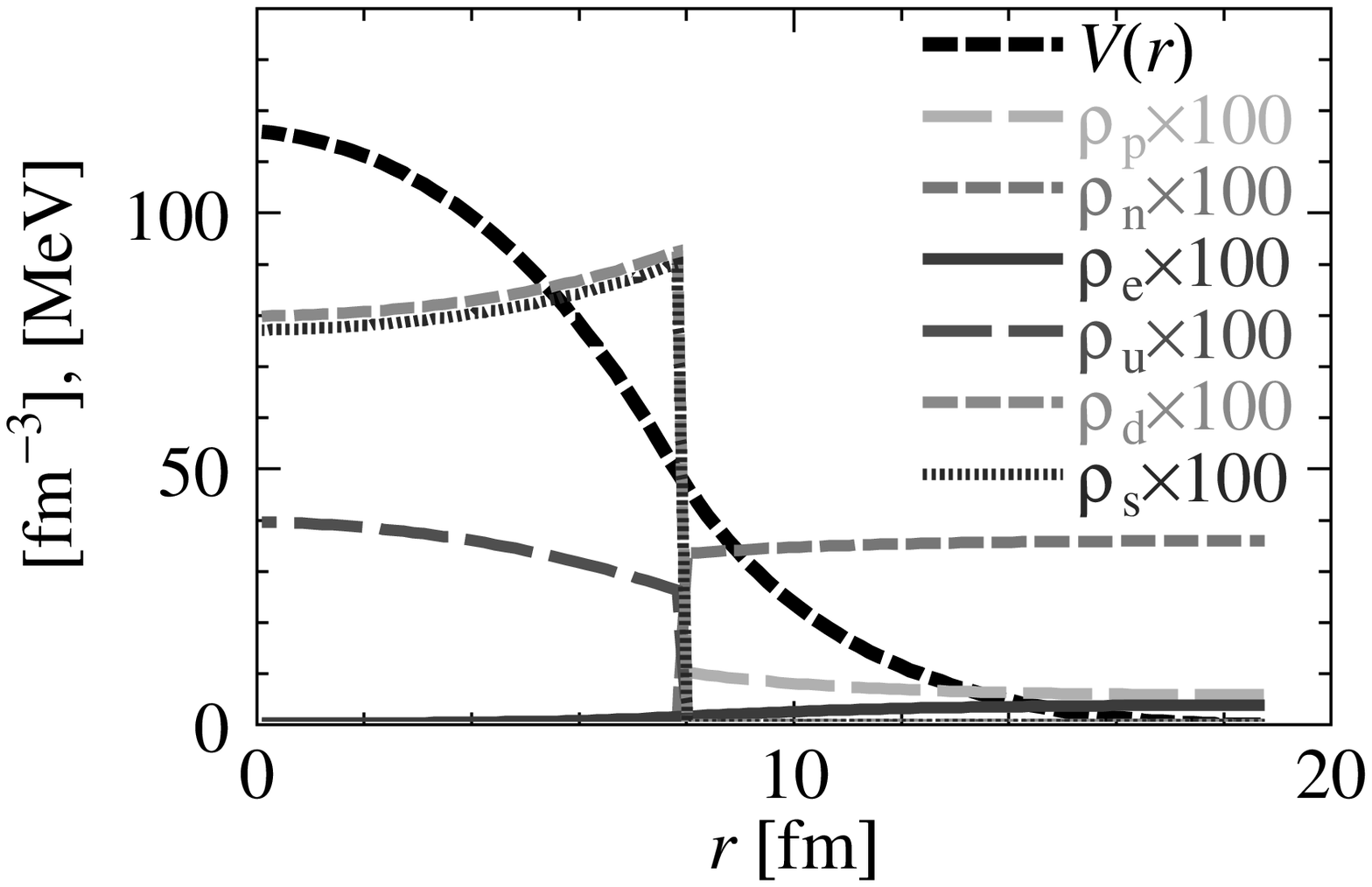}
\caption{Density profiles and the Coulomb potential given by the self-consistent
 calculation for the same parameter set as Fig.~5. $R=7.7$ fm and
 $R_W=18.9$ fm.}
\label{densprof-sc}
%\end{tabular}
\end{minipage}
%\end{center}
\end{figure}
%\end{wrapfigure}
%
If the surface tension is stronger, the relative importance of the
screening effect becomes smaller and the effect of the surface tension
becomes more dominant, as is seen in Fig.~4.
\begin{wrapfigure}{r}{70mm}
%\begin{figure}[htb]
%\begin{tabular}{c c}
%\begin{center}
\includegraphics[width=70mm]{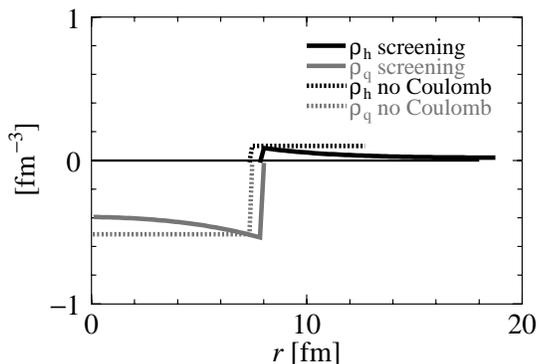}
%\end{tabular}
\caption{ Local-charge densities for the ``no Coulomb'' case and the
 case of the self-consistent with the screening effect. For ``no Coulomb'', charge
 density of each phase is constant over the region. The absolute
 value of the charge density is
 larger than that given by the self-consistent calculation in each phase. 
 In the hadron phase, the charge density becomes almost 
 vanished near the cell boundary $r=R_W$.}
\label{chdensprof}
%\end{center}
%\end{figure}
\end{wrapfigure}
To be more realistic we have to take into
account the modification of the surface tension as the structure size changes. 
Though we cannot clearly say how the surface tension is affected, 
we may infer from the previous study that it is not so large.   
Although the charge screening has not so large effects on bulk properties of the matter, 
we shall see that it is remarkable for the charged particles to change the properties of the mixed phase. 

The screening effect induces
the rearrangement of the charged particles. 
We can see this screening effect by comparing 
Fig.\ \ref{densprof-no} with Fig.\ \ref{densprof-sc}.
The quark
phase is negatively charged and the hadron phase is positively charged. 
The negatively charged particles in the quark phase such as {\it d},
{\it s}, {\it e}  and the positively charged particle in the hadron phase {\it p} are
attracted toward the boundary.
On the contrary the positively charged particle in the quark phase {\it u}
and  negatively charged particle in the hadron phase {\it e} are repelled
from the boundary.
The charge screening effect also reduces the net charge in each phase.
In Fig.\ \ref{chdensprof}, we show the local charge densities
of the two cases shown in Figs.\ \ref{densprof-no} and \ref{densprof-sc}.
The change of the number of charged particles due to the screening is as follows:
In the quark phase, the numbers of $d$ and $s$ quarks and
electrons decrease,
while the number of $u$ quark increases. 
In the hadron phase, on the other hand, 
the proton number should decrease and the electron number
should increase. 
Consequently the local charge decreases in the both phases.
In Fig.\ \ref{chdensprof} we can see that the core region of the droplet
tends to be charge-neutral and 
 near the boundary of the Wigner-Seitz cell 
is almost charge-neutral.

\begin{figure}[htb]
\begin{minipage}[t]{70mm}
%\includegraphics[width=8cm]{edens_rho.eps}
%\begin{center}
\includegraphics[width=70mm]{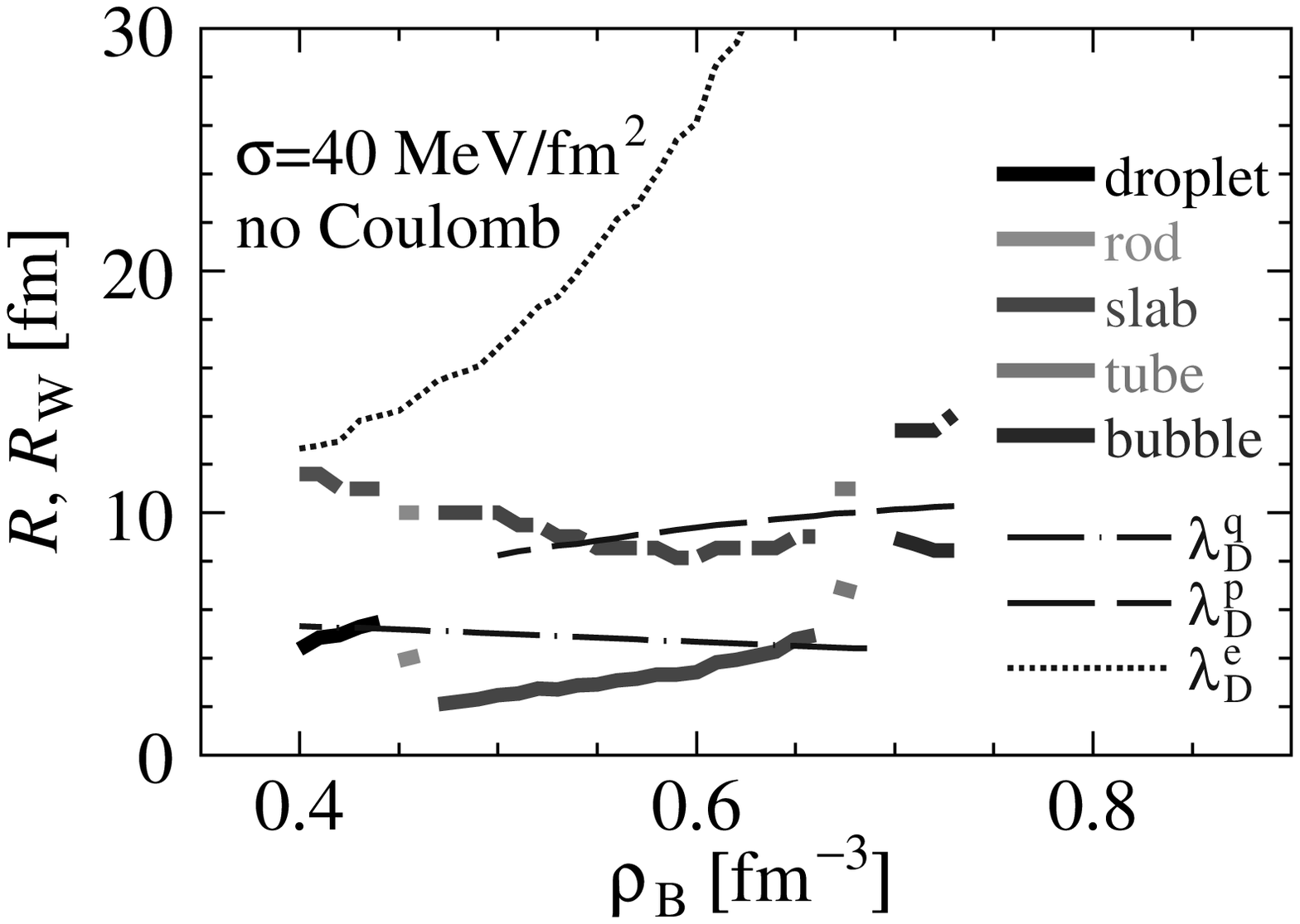}
\caption{ Lump and cell radii given by the ``no Coulomb'' 
 calculation. The Debye screening length is also depicted for
 comparison. $R$ is thick-solid line and $R_W$ is thick-dashed line. 
We can see the size of the structure becomes less than the
 Debye screening length.}
\label{rho-cell40no}
\end{minipage}
\hspace{8pt}
\begin{minipage}[t]{70mm}
\includegraphics[width=70mm]{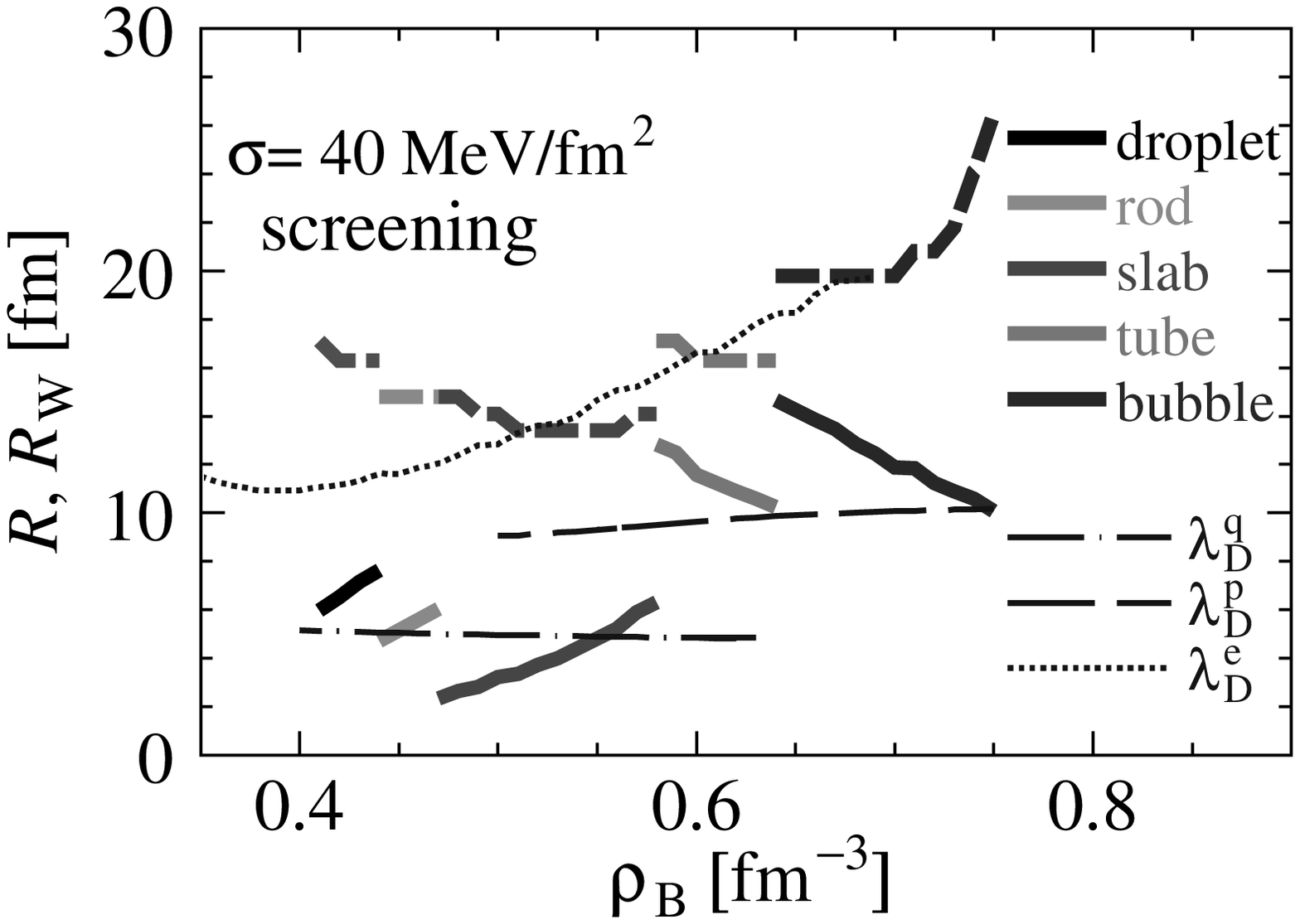}
\caption{ Same as Fig.~8 given by the self-consistent calculation with the
 screening effect. The size of the structure becomes larger than 
that given by ``no Coulomb'', and consequently exceeds the 
 Debye screening length.}
\label{rho-cell40sc}
\end{minipage}
%\end{center}
\end{figure}
%\end{wrapfigure}
%
\begin{wrapfigure}{r}{80mm}
%  \epsfxsize = \halftext
%  \centerline{ \epsfbox{edens_rho.eps}}
%\begin{center}
%\includegraphics[width=40mm]{lambda_no.eps2}
%\caption{ schematic image between droplet size and screening length.}
%\label{lambda_sc}
%\end{minipage}
%\begin{minipage}[t]{40mm}
\includegraphics[width=80mm]{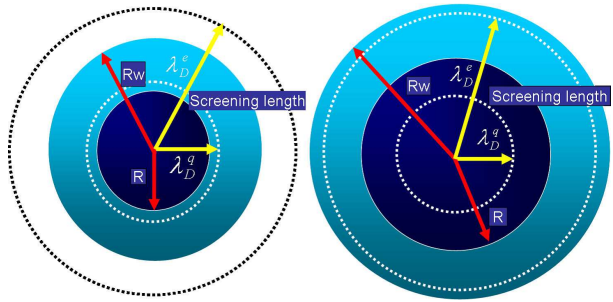}
\caption{Schematic graphs of the droplet size and the Debye screening
 length. Right figure shows the case of the self-consistent calculation with the
 screening effect and left figure ``no Coulomb''.}
%\label{lambda_no}
\label{lambdadrop}
%\end{minipage}
%\end{center}
%\end{figure}
\end{wrapfigure}
In Figs.~\ \ref{rho-cell40no} and \ref{rho-cell40sc} we present the lump
and cell radii for each density.
As we have shown in the previous paper\cite{vos}, the Coulomb energy
is suppressed for larger $R$ by the screening effect. 
The $R$ dependence of the total thermodynamic potential comes from the
contributions of the surface tension and the Coulomb interaction: 
the optimal radius giving the minimum of the thermodynamic potential 
is then determined by the balance between two
contributions, since the
former gives a decreasing function, while the latter an increasing one.
If the Coulomb energy is
suppressed, the minimum of the thermodynamic potential is shifted
to larger radius. 
As a result the size of the embedded phase ($R$) and the cell size ($R_W$) become large.
In Ref.~\cite{vos} they demonstrated that the minimum disappears for
a large value of the surface tension parameter: the structure becomes 
mechanically unstable in this case. 
We cannot show it directly in our framework because such unstable solutions 
are automatically excluded during the numerical procedure, 
while we can see its tendency 
in Figs.\ \ref{rho-cell40no} and \ref{rho-cell40sc}:
$R$ and $R_W$ get larger by the screening effect.

We also see the relation between the size of the geometrical structure and the Debye
screening length.
The Debye screening length appears in the {\it linearized} Poisson
equation 
and is then given as 
\begin{equation}
\left(\lambda^{q}_D\right)^{-2}\!\!=\! 4 \pi \sum_f Q_f \! \left( \frac{\partial \langle
				     \rho_f^\mathrm{ch}
				     \rangle}{\partial \mu_f} \right), \hspace{5pt}
\left(\lambda^{p}_D\right)^{-2}\!\!=\! 4 \pi Q_p \! \left( \frac{\partial \langle
				     \rho_p^\mathrm{ch}
				     \rangle}{\partial \mu_p} \right), \hspace{5pt}
\left(\lambda^{e}_D\right)^{-2}\!\!=\! 4 \pi Q_e \! \left( \frac{\partial \langle
				     \rho_e^\mathrm{ch}
				     \rangle}{\partial \mu_e} \right), 
\label{lambda}
\end{equation}
where $\langle \rho_f^\mathrm{ch} \rangle$ stands for 
the averaged density in quark phase, $\langle \rho_p^\mathrm{ch}
\rangle$ is proton number averaged density in the hadron phase 
and $\langle \rho_e^\mathrm{ch} \rangle$
is the electron charge density averaged inside the cell.
It gives a rough measure for the screening effect:
At a distance larger than the Debye screening length, 
the Coulomb interaction is effectively suppressed. 

\begin{figure}[htb]
\begin{minipage}[t]{70mm}
%\includegraphics[width=8cm]{edens_rho.eps}
%\begin{center}
\includegraphics[width=70mm]{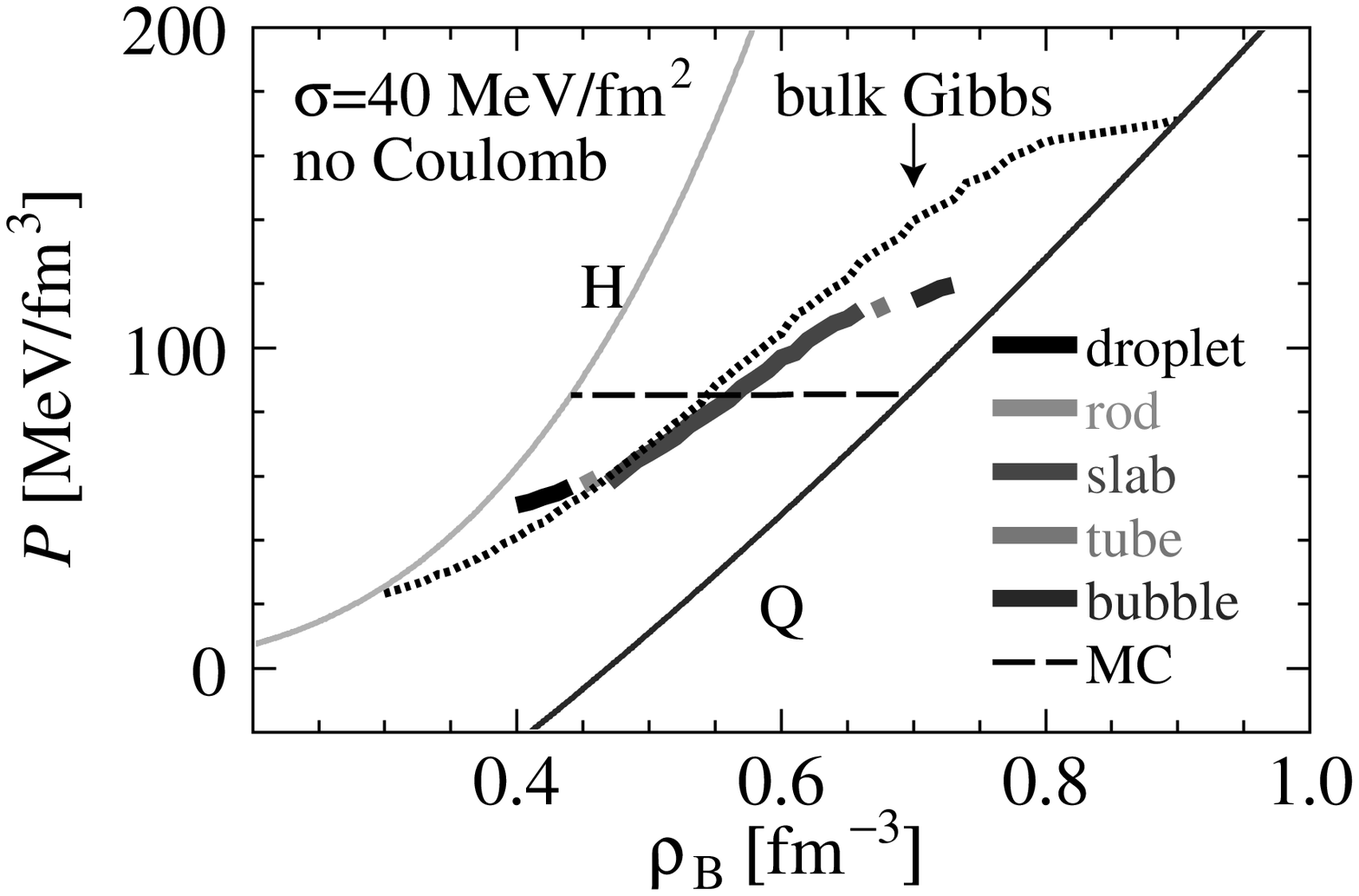}
\caption{ Pressure as a function of baryon-number density given by the
 ``no Coulomb'' calculation for $\sigma=40$MeV/fm$^2$. The results given
 by ``bulk Gibbs'' and MC are also presented for comparison. }
\label{pres40no}
\end{minipage}
\hspace{8pt}
\begin{minipage}[t]{70mm}
\includegraphics[width=70mm]{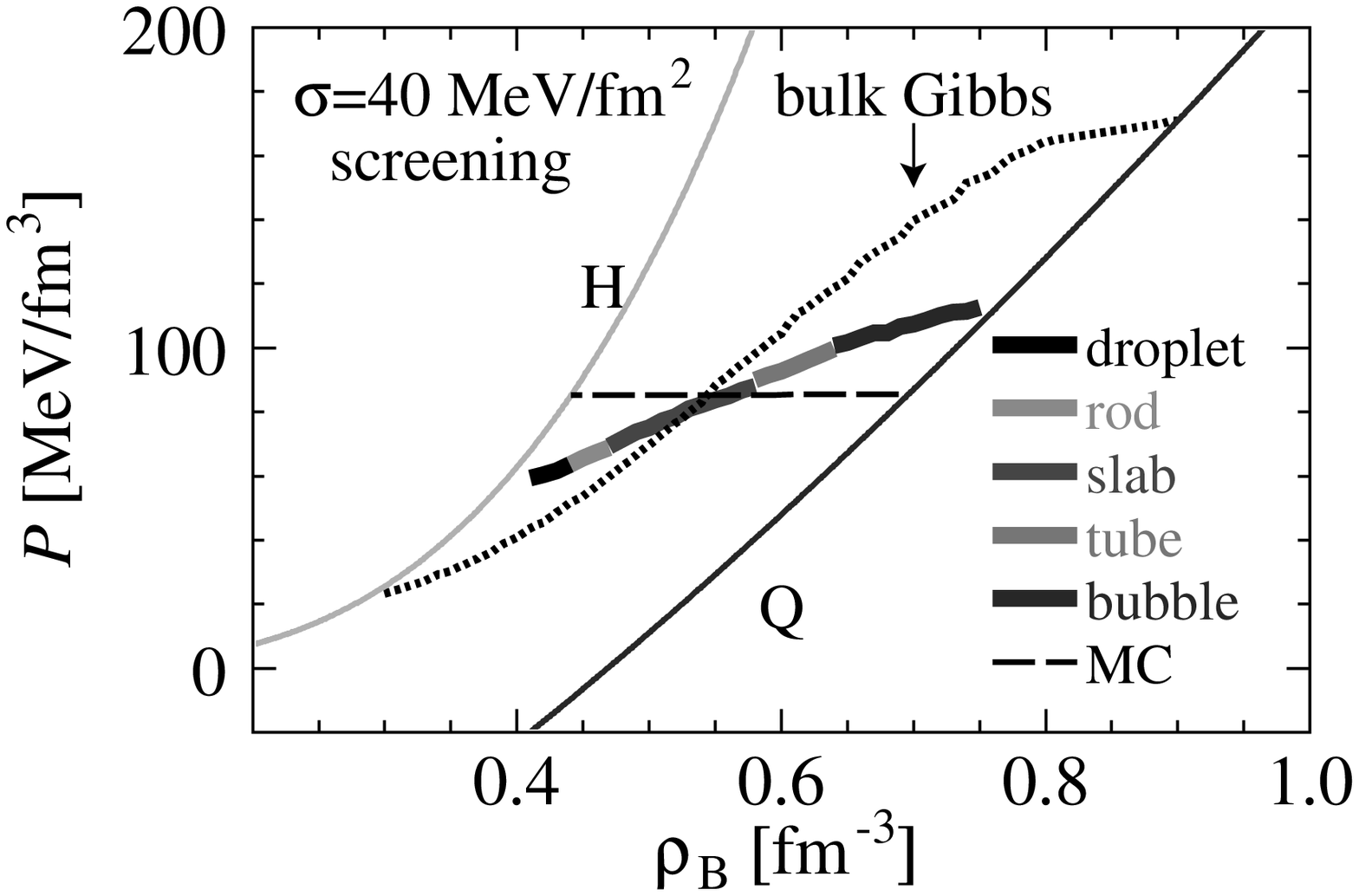}
\caption{  Same as Fig.~\ref{pres40no} given by the self-consistent calculation with the
 screening effect.}
\label{pres40sc}
\end{minipage}
%\end{center}
\end{figure}

In Fig.\ \ref{rho-cell40no} we show sizes of geometrical structure
for ``no Coulomb'' case.
If we ignore the screening effect, the size of the embedded phase is
comparable or smaller than the corresponding quark Debye screening
length $\lambda_D^q$ (Fig.\ \ref{lambdadrop}). 
This may mean that the Debye
screening is not so important.
Actually, many authors have neglected the screening
effect due to this argument\cite{pet,alf2}. 
In Fig.\ \ref{rho-cell40sc}, however, 
we see that the size of the embedded phase 
can be larger than $\lambda_D^q$ (Fig.\ \ref{lambdadrop})
in the self-consistent calculation.
We can also see the similar situation about $R_W$ and $\lambda_D^e$.
This means that the screening has important
effects in this mixed phase. 
We cannot expect such a effect
without solving the Poisson equation because of the non-linearity.
We show the EOS 
in Figs.\ \ref{pres40no} and \ref{pres40sc}. 
The pressure of the mixed phase becomes similar to that given
by MC due to the screening effect.
%\end{wrapfigure}

We have used 
 a fixed surface tension parameter in the present study.
Surface tension is a very difficult problem because it should be
self-consistent with the two phases of matter, quark and hadron.
Lattice QCD, based on the first principle, would be the most reliable
theory.
It predicts that the surface tension can be
10-100 MeV/fm$^2$\cite{kaja,huan}.
Although this range is for high temperature,
our choice is  within it.
Moreover, other model calculations of the surface tension\cite{mad1,mad2,berg,mond} are similar to our
choice.
Although we cannot conclude that MC is {\it perfectly} correct, 
we can say that the results obtained by the ``no Coulomb'' calculation,
which many authors have used,
have to be checked again 
taking into account the finite-size effects.

Let us consider some implication of these results for neutron star phenomena.
Glendenning\cite{gle2} suggested many SMP appear
in the core region by using ``bulk Gibbs'':
the mixed phase should appear for several kilometers. 
However we can say that the region of SMP
should be narrow in the $\mu_\mathrm{B}$ space and EOS is more similar
to that of MC 
due to the finite-size effects.
These results correspond to recent other calculations.
Bejger et al.\ \cite{bejg} have
examined the relation between the mixed phase
and glitch phenomena, and shown that the mixed phase should be narrow if the
glitch is generated by 
the mixed phase in the inner core. 
On the other hand the gravitational
wave asks for density discontinuity in the core region\cite{mini}. 
These studies support our result.

\section{Summary and concluding remarks}\label{summary}

We have numerically studied the charge screening effect in the hadron-quark mixed phase, 
by fully including the non-linear effects in the Poisson equation.
Comparing the results with those given by ``no Coulomb'' calculation, we have
elucidated the screening effect. The density profiles of the charged
particles are much modified by the screening effect, while the
thermodynamic potential is not so much affected; the charge
rearrangement induced by the screening effect tends to make the net
charge smaller in each phase. Consequently the system tends to be
locally charge-neutral, which suggests 
that MC is effectively justified even if it is thermodynamically incorrect.
In this context, it would be interesting to refer to the work by 
Heiselberg \cite{hei}, who  studied the screening effect on a quark
droplet (strangelet) in the vacuum,
and suggested the importance of the rearrangement of charged particles.

We have seen that thermodynamic quantities such as thermodynamic potential and
pressure become close to those derived from MC by the screening effect,
which also suggests that MC is effectively justified due to the
screening effect. 
As another case of more than one chemical potential system, kaon condensation  
has been also studied\cite{maru1} and the results are similar to those
in 
the present study. Thus the importance of the screening effect should be
a common feature for the first-order phase transitions in high-density matter.

We have included the surface tension at the hadron-quark interface, 
while its definite value is not clear at present. 
There are also many
estimations for the surface tension at the hadron-quark
interface in lattice QCD \cite{kaja,huan}, in shell-model
calculations \cite{mad1,mad2,berg} and in model
calculations based on the dual-Ginzburg Landau theory \cite{mond}.
Our parameter is in that reasonable range.

We have considered some implications of our results for neutron star phenomena.
The screening effect would restrict
the allowed SMP region in neutron stars, in contrast with a wide region
given by ``bulk Gibbs''\cite {gle2,gle3}. It could be said that
they should change the bulk 
property of neutron stars, especially the structure of the core region. 

Compact stars
have the strong magnetic field and its origin is not well understood. One
possibility is that it comes from quark matter in the
core\cite{tat1,tat2,tat3}. Therefore it should be interesting to 
include the magnetic field contribution in our formalism.

We have assumed zero temperature here. 
It would be  much interesting to include the finite-temperature
effect. Then it is possible to draw the phase diagram in the
$\mu_\mathrm{B}$ - $T$ plane and we can study the properties of the
deconfinement phase transition; our study may be extended to treat 
the mixed phase to appear during the hadronization of QGP in the
nucleus-nucleus collisions 
and supernova explosions.

 In this study we have used a simple model for quark matter to
figure out the finite-size effects in the SMP. However,
it has been suggested that the color superconductivity would be a ground state of quark
matter \cite{alf1,alf2}. Hence we will include it in a further study.
The hadron phase should be also treated more realistically; for example, 
we should include the
hyperons or kaons in hadron matter.
In the recent studies the
mixed phase has been also studied \cite{shov,redd} in the context of
various phases in the color superconducting phase.

\section*{Acknowledgements}
We wish to acknowledge Dr. D.N. Voskresensky and Dr. T. Tanigawa for fruitful discussion. 
T.E. appreciates Dr. H. Suganuma for his encouragement.
This work is supported in part by the Grant-in-Aid for the 21st Century
COE ``Center for the Diversity and Universality in Physics'' from the
Ministry of Education, Culture, Sports, Science and Technology of Japan.
The work of one of the authors (T.T.) is partially supported by 
the Japanese Grant-in-Aid for Scientific
Research Fund of the Ministry of Education, Culture, Sports, Science and
Technology (13640282,16540246).

\end{document}